# A Lower Bound for Estimating High Moments of a Data Stream

Sumit Ganguly

Indian Institute of Technology, Kanpur

**Abstract**

We show an improved lower bound for the $F_p$ estimation problem in a data stream setting for $p > 2$. A data stream is a sequence of items from the domain $[n]$ with possible repetitions. The frequency vector $x$ is an $n$-dimensional non-negative integer vector $x$ such that $x(i)$ is the number of occurrences of $i$ in the sequence. Given an accuracy parameter $\Omega(n^{-1/p}) < \epsilon < 1$, the problem of estimating the $p$th moment of frequency is to estimate $\|x\|_p^p = \sum_{i \in [n]} |x(i)|^p$ correctly to within a relative accuracy of $1 \pm \epsilon$ with high constant probability in an online fashion and using as little space as possible. The current lower bound for space for this problem is $\Omega\big(n^{1-2/p}\epsilon^{-2/p} + n^{1-2/p}\epsilon^{-4/p}/\log^{O(1)}(n) + (\epsilon^{-2} + \log(n))\big)$. The first term in the lower bound expression was proved in [2, 3], the second in [6] and the third in [5]. In this note, we show an $\Omega(p^2 n^{1-2/p} \epsilon^{-2}/\log(n))$ bits space bound, for $\Omega(pn^{-1/p}) \le \epsilon \le 1/10$.

## 1  Introduction

In the insert-only data streaming model, a stream is modeled as a sequence of items $i_1, i_2, \ldots$, where the items come from a large domain $[n] = \{1, 2, \ldots, n\}$. The frequency vector is an $n$-dimensional vector $x$ whose $i$th coordinate $x(i)$ counts the number of occurrences of $i$ in the sequence. Each new arrival of an item $i_j$ increments $x(i_j)$ to $x(i_j) + 1$. Define $\|x\|_p^p = \sum_{i \in [n]} |x_i|^p$. The $p$th moment estimation problem, with accuracy parameter $\epsilon$, is to design a structure that can process the stream sequence in an online fashion and return a real value $\hat{F}_p$ satisfying $\big|\hat{F}_p - \|x\|_p^p\big| \le \epsilon \|x\|_p^p$ with probability $9/10$. The estimate $\hat{F}_p$ may use only the structure and not the original stream, that is, a stream may be processed in an online fashion only. The $F_p$ estimation problem has played a pivotal role in the study of data streaming algorithms. It was first posed and studied by Alon, Matias and Szegedy [1]. They showed that for all $p \ne 1$, a deterministic $\epsilon$-accurate $F_p$ estimation with $\epsilon \le 1/8$ requires $\Omega(n)$ bits, as does a randomized algorithm with no error. This reduces the scope to approximate randomized algorithms or randomized PTAS. A series of works [1, 2, 3] culminated in showing a lower bound of $\Omega(n^{1-2/p}\epsilon^{-2/p})$ bits for $\epsilon$-accurate $F_p$ estimation. Very recently, Woodruff and Zhang in [6] improve this bound to $\tilde{\Omega}(n^{1-2/p}\epsilon^{-4/p})$ bits, where, $\tilde{\Omega}(f(n,\epsilon))$ denotes $f(n,\epsilon)/\log^{O(1)}(n/\epsilon)$. Woodruff in [5] shows an $\Omega(\epsilon^{-2} + \log(n))$ bits bound for $F_p$, for all $p \ne 1$. [1] So, the current lower bound for $F_p$ estimation in bits is:

$$\Omega\Big(n^{1-2/p}\epsilon^{-2/p} + \frac{n^{1-2/p}\epsilon^{-4/p}}{\log^{O(1)} n} + \epsilon^{-2} + \log(n)\Big)$$

In this note, we show a lower bound of $\Omega\big(p^2 n^{1-2/p}\epsilon^{-2}/\log(n)\big)$ bits for this problem, improving upon the current known bounds.

---

[1] Jayram and Woodruff show $\Omega(\epsilon^{-2}\log(n))$ bits bound when deletions are also allowed, for all $p \ge 0$.

## 2 Lower Bound

We will reduce the standard $t$-party set disjointness problem to $F_p$ estimation. The problem $t$-DISJ is as follows: the instance is a collectionof $t$ sets $S_1, \ldots, S_t$, each subset of $[n]$, where, the set $S_i$ is given to the $i$th party with the promise that the set family is either pair-wise disjoint, or, $S_1 \cap \ldots \cap S_t$ has exactly one element in common. We denote the $i$th coordinate of a vector $x$ by $x(i)$; so $x = [x(1), \ldots, x(n)]$. With this notation, an instance of $t$-DISJ consists of $n$-dimensional binary vectors $x_1, \ldots, x_t$, where, $x_r$ is given to the $r$th party and is interpreted as the characteristic vector of the set $S_r$. The promise is that either, (a) $x_1 + \ldots + x_t$ is a binary vector (the disjoint case), or, (2) there is exactly one index $i$ such that $x_1(i) = x_2(i) = \ldots = x_t(i) = 1$ (the common element case). It is well-known that any one-way randomized communication protocol that solves $t$-DISJ with probability at least $7/8$ requires $\Omega(n/t)$ bits [2, 3]. We show the following theorem.

**Theorem 1** *For $2 < p < n^{1/p}/2$ and $\max(80p/n^{1/p}, 3/\sqrt{n}) \le \epsilon \le 1/4$, an algorithm that estimates $F_p$ with relative error of $\epsilon/10$ and with probability $19/20$ uses space $\Omega\bigl(\frac{p^2 n^{1-2/p}}{\epsilon^2 \log(n)}\bigr)$ bits.*

**Proof** We present a randomized one-way communication protocol for $t$-DISJ that is correct with probability $9/10$, where, $t = \lceil \epsilon n^{1/p}/(2p) \rceil$. The protocol uses two structures that can process stream updates, one for estimating $F_p$ to within a factor of $1 \pm \epsilon/10$ with confidence $1 - 1/(20n)$, and, the second for estimating $F_0$ to within a factor of $1 \pm \epsilon/10$ with probability $19/20$.

A one-way protocol for $t$-DISJ is as follows. Consider an instance of $t$-DISJ. Party 1 inserts $x_1$ into each of the structures for estimating $F_p$ and $F_0$ and sends the pair of structures to the second party. This party further adds its vector $x_2$ into the two structures received and then relays it to the third party, and so on, in sequence. Finally, the $t$th party inserts its own vector into the structures obtained from $t-1$st party. It then uses the procedure InferDisj of Figure 1 to infer whether the instance is pair-wise disjoint or has a common element.

We first show that the procedure InferDisj is correct with probability at least $9/10$. Define the event $\text{GOODF}_0$ as $\hat{F}_0 \in (1 \pm \epsilon/10)\|x\|_0$, so, $\text{GOODF}_0$ holds with probability $19/20$. Let $x = x_1 + x_2 + \ldots + x_t$. Say that $i$ is a heavy item in $x$ if $x(i) = t$. Procedure INFERDISJ obtains an estimate $\hat{F}_p^i$ obtained by applying the $F_p$ estimation algorithm to the vector $x + n^{1/p} e_i$ (in parallel, for each $i$). Given $x$ and an index $i$, we consider three cases. Assume $3p < n$.

**Case 1:** $x$ has no heavy item, that is, $x$ is a binary vector. So, $x + n^{1/p} e_i = x' + (n^{1/p} + x(i)) e_i$, where, $x'$ is a binary vector with $x'(i) = 0$. Hence, $\|x'\|_0 = \|x\|_0 - x(i)$ and

$$\begin{aligned}
\|x + n^{1/p} e_i\|_p^p &= \|x'\|_0 + (n^{1/p} + x(i))^p \\
&\le \|x\|_0 + n e^{x(i) p / n^{1/p}} - x(i) \\
&\le \|x\|_0 + n(1 + 5p/(4n^{1/p})), \quad \text{assuming } p < n^{1/p}/3 \text{ and elementary calculations} \ . \\
&\le \|x\|_0 + n(1 + \epsilon/64), \quad \text{since, } 5p/(4n^{1/p}) \le \epsilon/64.
\end{aligned} \quad (1)$$

So with probability $1 - 1/(20n)$, and conditional on $\text{GOODF}_0$,

$$\begin{aligned}
\hat{F}_p^i &\le (1 + \epsilon/10)\|x + n^{1/p} e_i\|_p^p \\
&\le (1 + \epsilon/10)(\|x\|_0 + n(1 + \epsilon/64)), \quad \text{from (1)} \\
&< \hat{F}_0 + n(1 + \epsilon/8) \ .
\end{aligned} \quad (2)$$

*procedure* InferDisj
*Input*: Given $F_0$ and $F_p$ sketches of $x = x_1 + \ldots + x_t$ (integer $n$-dimensional vector) such that

1. For $x$, one of the two cases hold: *Disjoint* : $x \in \{0,1\}^n$, or, *Common Element*: there exists exactly one $i$ such $x(i) = t = \lceil \epsilon n^{1/p}/(2p) \rceil$ and the remaining $x(j)$'s are either 0 or 1.

2. $\hat{F}_0 \in (1 \pm \epsilon)F_0$ with probability $19/20$, and, $\hat{F}_p \in (1 \pm \epsilon)F_p$ with probability $1 - 1/(20n)$.

*Output*: Returns COMMON ELEMENT $i$ if the input is identified to fall in the *Common Element* case and the item with frequency $t$ is identified as $i$, and, returns DISJOINT if the input is identified to fall in the *Disjoint* case.

1.    $\hat{F}_0 = $ Estimate for $\|x\|_0$.
2.    **for** $i := 1$ to $n$ **in parallel do** {
3.        **insert** $(i, n^{1/p})$ to $F_p$ sketch
4.        Obtain $\hat{F}_p$
5.        **if** $\hat{F}_p \geq \hat{F}_0 + n(1 + 2\epsilon/5)$ **then**
6.           **return** COMMON ELEMENT $i$
7.    }
8.    **return** DISJOINT

Figure 1: Solving Set Disjointness using $F_p$ and $F_0$ sketches

**Case 2:** $x$ has a (unique) heavy item whose index is $j \neq i$. Then, $x + n^{1/p}e_i = x' + te_j + (n^{1/p} + x(i))e_i$, where, $x'(i) = x'(j) = 0$. Hence, $\|x'\|_0 = \|x\|_0 - 1 - x(i)$ and

$$\|x + n^{1/p}e_i\|_p^p = \|x\|_0 - 1 - x(i) + t^p + (n^{1/p} + x(i))^p$$

$$\leq \|x\|_0 - 1 + \frac{\epsilon^p n}{(2p)^p} + ne^{x(i)p/n^{1/p}} - x(i)$$

$$\leq \|x\|_0 - 1 + \frac{\epsilon n}{4^{2p-1}} + n(1 + 5p/(4n^{1/p})) - 1, \text{ as in } (1).$$

$$\leq \|x\|_0 + n\left(1 + \frac{\epsilon}{64} + \frac{\epsilon}{64}\right) \tag{3}$$

In the second to last step, we use $\epsilon^p \leq \epsilon(1/4)^{p-1}$ and $(2p)^p \geq 4^p$ since $p \geq 2$. In the last step, we make use of the assumption that $\frac{5p}{4n^{1/p}} \leq \frac{\epsilon}{64}$. Hence, with probability $1 - 1/(20n)$, and conditional on GOODF$_0$,

$$\hat{F}_p^i \leq (1 + \epsilon/10)\|x + n^{1/p}e_i\|_p^p$$
$$\leq (1 + \epsilon/10)(\|x\|_0 + n(1 + \epsilon/32)), \quad \text{from } (3)$$
$$\leq (1 + \epsilon/10)\|x\|_0 + n(1 + 2\epsilon/17), \quad \text{using } \epsilon \leq 1/4$$
$$\leq \hat{F}_0 + n(1 + \epsilon/7) \tag{4}$$

**Case 3:** $x$ has a (unique) heavy item with index $i$. Then, $x + n^{1/p}e_i = x' + (n^{1/p} + \epsilon n^{1/p}/(2p))e_i$, where, $x'$ is a binary vector with $x'(i) = 0$. Hence, $\|x'\|_0 = \|x\|_0 - 1$ and

$$\|x + n^{1/p}e_i\|_p^p = \|x'\|_0 + (n^{1/p} + t)^p = \|x\|_0 - 1 + n\left(1 + \frac{\epsilon}{2p}\right)^p \geq \|x\|_0 + n(1 + \epsilon/2) \tag{5}$$

The last step uses a two-term Taylor expansion of $(1+\alpha)^p$ around $\alpha = 0$ to obtain, $(1+\alpha)^p \geq 1 + p\alpha + p(p-1)\alpha^2/2$. Setting $\alpha = \frac{\epsilon}{2p}$, we get $(1+\alpha)^p \geq 1 + \frac{\epsilon}{2} + \frac{p^2\epsilon^2}{8p^2} - \frac{p\epsilon^2}{8p^2} \geq 1 + \frac{\epsilon}{2} + \frac{15\epsilon^2}{32}$, since, $p \geq 2$. So, $n(1 + \frac{\epsilon}{2p})^p - 1 \geq n(1 + \frac{\epsilon}{2})$, since, $\epsilon \geq 3/\sqrt{n}$.

The procedure INFERDISJ estimates $\|x + n^{1/p}e_i\|_p^p$ using the assumed $F_p$ estimation procedure. So with probability $1 - 1/(20n)$, and conditional on $\text{GOODF}_0$,

$$\hat{F}_p^i \geq (1 - \epsilon/10)\|x + n^{1/p}e_i\|_p^p \geq (1 - \epsilon/10)(\|x\|_0 + n(1 + \epsilon/2)) \geq \hat{F}_0 + n(1 + 2\epsilon/5), \quad (6)$$

Define the event $\text{GOODF}_p$ to hold if $\hat{F}_p^i \in (1 \pm \epsilon/10)\|x + n^{1/p}e_i\|_p^p$, for each $i \in [n]$. By union bound, $\text{GOODF}_p$ holds with probability $19/20$. Similarly, $\text{GOODF}_0$ holds with probability $19/20$. Hence, both $\text{GOODF}_p$ and $\text{GOODF}_0$ hold simultaneously with probability $1 - 2/20 + 1/400 > 9/10$. Assume $\text{GOODF}_p$ and $\text{GOODF}_0$ hold. Then, procedure INFERDISJ is correctly able to distinguish Case 3 from Cases 1 or 2, since for Case 3, $\hat{F}_p \geq \hat{F}_0 + n(1 + 2\epsilon/5)$ and for Cases 1 and 2, $\hat{F}_p \leq \hat{F}_0 + n(1 + \epsilon/7)$. Case 3 corresponds to the common element case when the common element is $i$. Case 2 corresponds to the common element case but the common element is not $i$. Finally Case 1 corresponds to the pair-wise disjoint sets case. (Cases 1 and 2 cannot be distinguished by the algorithm). So, assuming $\text{GOODF}_0$ and $\text{GOODF}_p$, the check for $\hat{F}_p^i \geq \hat{F}_0 + n(1 + 2\epsilon/5)$ in parallel succeeds only when the sets have $i$ as the common element. The check fails for all other values of $i$. If the sets are pair-wise disjoint, then, the check fails again. Since both $\text{GOODF}_0$ and $\text{GOODF}_p$ hold with probability $9/10$, it follows that procedure INFERDISJ solves $t$-DISJ with probability $9/10$.

By the work of [2, 3], any protocol for solving $t$-DISJ requires a total communication of $\Omega(n/t)$ bits. Let $S(\epsilon)$ be the total space used by the protocol proposed above. Then,

$$S(\epsilon)t = \Omega(n/t), \quad \text{or, } S(\epsilon) = \Omega(n/t^2) = \Omega(p^2 n^{1-2/p}/\epsilon^2)$$

The space $S(\epsilon) = S_0(\epsilon) + S_p(\epsilon)$, where, $S_0$ is the space required for a $(1 \pm \epsilon/10)$ approximation of $F_0$ with high constant confidence and $S_p(\epsilon)$ is the space required for a $(1 \pm \epsilon/10)$-approximation of $F_p$ with confidence $1 - 1/(20n)$. Since the above protocol does not involve deletions, from [4], $S_0(\epsilon) = O(\epsilon^{-2} + \log n)$ bits. Hence, $S(\epsilon) = S_p(\epsilon) + S_0(\epsilon) \geq \Omega(p^2 n^{1-2/p}\epsilon^{-2}))$, or,

$$S_p(\epsilon) \geq \Omega(p^2 n^{1-2/p}\epsilon^{-2})) - S_0(\epsilon)$$
$$= \Omega(p^2 n^{1-2/p}\epsilon^{-2}) - O(\epsilon^{-2} + \log n)$$
$$= \Omega(p^2 n^{1-2/p}\epsilon^{-2} - \log(n))$$

Since $S_p(\epsilon)$ is the space used for estimating $F_p$ to within $1 \pm \epsilon/10$ with confidence $1 - 1/(20n)$, it follows that the space required for estimating $F_p$ to within $1 \pm \epsilon/10$ with confidence $1 - 1/20$ is lower bounded by

$$\Omega\Big(\frac{S_p(\epsilon)}{\log(n)}\Big) = \Omega\Big(\frac{p^2 n^{1-2/p}}{\epsilon^2 \log(n)} - 1\Big) = \Omega\Big(\frac{p^2 n^{1-2/p}}{\epsilon^2 \log(n)}\Big)$$

where the last equality follows since there is an $\Omega(\epsilon^{-2} + \log(n))$ bound for the problem. ∎